# Effectively Searching Maps in Web Documents

Qingzhao Tan[†]    Prasenjit Mitra[†‡]    C. Lee Giles[†‡]

[†]Computer Science and Engineering  [‡]Information Sciences and Technology
The Pennsylvania State University, University Park, PA 16802, USA
qtan@cse.psu.edu, {pmitra, giles}@ist.psu.edu

Maps are an important source of information in archaeology and other sciences. Users want to search for historical maps to determine recorded history of the political geography of regions at different eras, to find out where exactly archaeological artifacts were discovered, etc. Currently, they have to use a generic search engine and add the term map along with other keywords to search for maps. This crude method will generate a significant number of false positives that the user will need to cull through to get the desired results. To reduce their manual effort, we propose an automatic map identification, indexing, and retrieval system that enables users to search and retrieve maps appearing in a large corpus of digital documents using simple keyword queries. We identify features that can help in distinguishing maps from other figures in digital documents and show how a Support-Vector-Machine-based classifier can be used to identify maps. We propose map-level-metadata e.g., captions, references to the maps in text, etc. and document-level metadata, e.g., title, abstract, citations, how recent the publication is, etc. and show how they can be automatically extracted and indexed. Our novel ranking algorithm weights different metadata fields differently and also uses the document-level metadata to help rank retrieved maps. Empirical evaluations show which features should be selected and which metadata fields should be weighted more. We also demonstrate improved retrieval results in comparison to adaptations of existing methods for map retrieval. Our map search engine has been deployed in an online map-search system that is part of the Blind-Review digital library system.

## 1  Introduction

Maps have long been used as traditional storage and retrieval media for geographic data. Currently, we can see map systems on the Web, which readily convey geographic information for different geographical areas. For example, GoogleMap[1] and MapQuest[2] provide details of local roads and streets as well as the nearby business. Maps are becoming more widely used in a variety of information needs.

Particularly, maps are an important type of figures which are frequently used in digital documents, especially academic and scientific documents in the field of archeology and geography. In these documents, maps are widely adopted for providing related geographic information to readers and enabling them better understand the document content visually. For instance, an archeology paper which describes pottery excavated from West Virginia, would probably first show a map around the area from where the pottery was located. Therefore, users of an archeology/geography digital library may be interested in searching maps in documents. An instance of such query is "Find documents which contain a map of Chaco Canyon". As more digital documents containing maps become available on the Web, there is a growing demand for a Web search system that provides users with tools to retrieve documents based on the information available within documents and within the map itself.

Despite the importance of conveying information using maps, to our best knowledge, no digital library has paid attention to geographical information in maps that are contained in digital documents. Often, archaeologists and other scientists have a need to search for maps from digital documents. If the collection of documents is extremely large, users typically append the term "map" with other keywords to search and retrieve geographical maps. However, this method is labor-intensive because of a large number of false positives in the search results. Thus, a special-purpose search engine that users can use to directly search for maps using keywords and get high-quality results is necessary.

---

[1] http://maps.google.com
[2] http://www.mapquest.com

In this paper, we address this problem by integrating the map search functionality into a scientific digital library. Our proposed map search system is novel. First, it differs from online map systems because it considers not only maps but also the specific linkage between maps and other related documents. Second, it differs from traditional approaches in digital libraries because it separates maps from general figures and text in a digital document and provides effective algorithms for searching them.

To build a map search system, we first construct a set of metadata for each map in a digital document. This metadata includes map level metadata (e.g., a map's caption) as well as document level matadata (e.g., a document's title, abstract). We denote the document containing the map the *host document* of that map. We then provide methods and algorithms to achieve the following three tasks:

- *Map Metadata Extraction*. A set of map metadata are extracted from the digital document. We first use TET[3] to extract text from PDF files and then apply a set of heuristics rules to locate the metadata in the extracted text.
- *Map Identification*. Maps are identified from the other non-map images. Distinguishing maps from other figures is a challenging problem. We use supervised learning methods and a set of features which are extracted from the map metadata.
- *Map Indexing and Retrieval*. Regarding the map metadata as various fields for map indexing, a novel index and a new ranking function are developed for the map search system.

Extensive experiments have been conducted to provide support for our proposed system. In our work, we focus on PDF documents due to its popularity in digital libraries.

The rest of this paper is organized as follows. First, we discuss the existing work related to our paper. We then provide the architecture design of the map search system. We list in detail the set of metadata we used to describe each map in a document. Based on these metadata, we show the methodologies to identify, extract, and index maps in digital libraries. Experimental results are provided to validate our proposed system followed by conclusions and future work.

## 2 Related Work

In this section, we will discuss related work.

**Existing Related Systems**. Much work has been done in the field of map data processing. A system used for acquisition, storage, indexing, and retrieval of map images is described in [1]. The inputs to their system are raster images of maps. These images are then stored as a record containing map related information in a relational database. Advanced database techniques from the field of spatial databases are adopted to build indices. Queries are posed to the system using an SQL-like language. In [2], a G-Portal system is set up to identify, classify, and organize geo-spatial and geo-referenced content on the World Wide Web. A digital library can access these data by using a map-based interface. A legend-driven map interpretation system, MARCO (MAp Retrieval by COntent), is proposed in [3] to convert map images from their physical representation to their logical representation. The logical representation is then stored and used to build index. R. P. Futrelle has set up a diagram understanding system to understand diagrams in technical documents [4,5]. This is the first such system to fully parse a variety of actual diagrams drawn from the research literature. Digmap system[4] [6] is a geographic IR system based on the historical digitized maps.

Our work is different from those we described above. In their approaches, the information is dynamically extracted from either paper-based maps or digital maps on the Web. They consider maps in isolation, i.e., excluding other content in digital documents. In contrast, we extract maps from digital documents and utilize the "context" of the maps like captions and references in the text, as well as other context-based boosting factors depending on the digital documents, to get hints on understanding their contents. Our experiment results show that inclusion of these factors can significantly improve the map retrieval performance. The DIGMAP system takes into account some textual description of the maps. However, these descriptions are provided along with the maps. DIGMAP does not consider the problem of extracting maps and the related map metadata from documents, as well as the problem of distinguishing maps from other images in a document. Both of these problems are critical because a large number of maps are embedded in documents.

---

[3] http://www.pdflib.com/products/tet/
[4] http://code.google.com/p/digmap/

There is another online map search online system[5] [7], which was motivated in part by discussions with our team, in which maps are also extracted from some Web documents in PDF format The maps are indexed based on maps' content in three dimensions, region, time period, and theme. However, this project is just starting and no details of the design or implementation are discussed in [7].

**Multimodal/Structured Document Retrieval**. Our work is also related to indexing and retrieving images for multimodal digital documents. Multimodal documents convey information using both text and images. Different Information Retrieval (IR) techniques have been used to build up, index and retrieval functions. Some of these existing works deal with only the text in the document and ignore all the image-related information [8]. Some of them deal with only images, as in content-based image retrieval (CBIR) [9]. Others use both text and image objects [10]. Current OCR tools were not very effective in extracting text from maps in our preliminary experiments. In the future, we will extract and index text lying inside maps. Once such information is extracted, our current framework can index and utilize them seamlessly.

As discussed in Section 6, we consider the map metadata as various information fields for a map, and propose a structured document retrieval technique to build a map index. Therefore, we now discuss related work on structured document retrieval briefly. A document is said to be structured when it contains multiple fields. A document's field structure is commonly used to improve retrieval performance in practice. The most commonly used approach for structured document retrieval is a score/rank linear combination [11,12,13,14], which treats each field as a separate document and computes a combined scores/ranks. In computing scores for each field, any ranking function for unstructured document retrieval can be adopted. Another approach is to essentially combine term frequencies instead of scores. In [15], an extension of the BM25 formula is introduced and is a simple and efficient method which combines term frequencies instead of field scores. In [16], within the language model framework, this approach proposes a separate language model for each field and then combines them linearly. Our work adopts and extends these two pieces of work and proposes novel retrieval functions for the map search system.

## 3   System Architecture

In this section, we introduce the overall system architecture. Figure 1 shows a simplified diagrammatic view of the system architecture. Our map search system has the following four components.

– *Document Preprocessor*. A large set of electronic academic and scientific documents can be obtained by crawling various open-access digital libraries and homepages of researchers in archaeology and related areas. Because most of these electronic scientific documents are scanned documents, we rely on an OCR system to first transfer the files to regular PDF files. After that, both textual information and figures are extracted from the PDF files by the document preprocessor.
– *Map Metadata Extractor*. Two kinds of metadata are extracted from the documents: host document level metadata and figure level metadata.

---

[5] http://scilsresx.rutgers.edu/ gelern/maps/

- *Map Identifier.* The input to the map identifier is a combination of extracted figures and figure metadata. The map identifier classifies the figures into two categories: maps and non-maps. The output maps along with their metadata are then stored in a map database. The corresponding map metadata is indexed.
- *Map Retrieval Subsystem.* The retrieval subsystem consists of the index, ranking algorithm, and the user interface. We build the text index using the Lucene tool[6]. Both the map metadata and its host document metadata are indexed. Given a query, the ranking algorithm generates a ranking score for each map in the map database. The user interface allows users to input keyword queries and returns a ranked list of document snippets. An example of the snippet list returned from the input query "Mississippian" is shown in Figure 2.

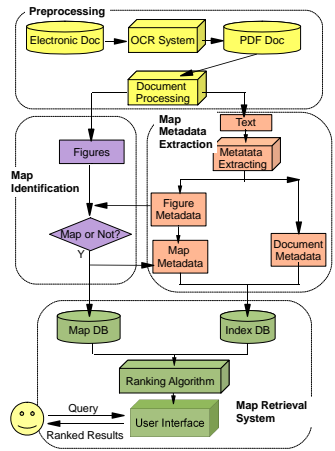

**Fig. 1.** The Architecture of a Map Search System.

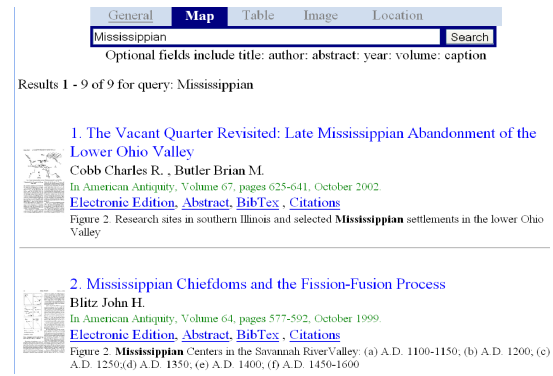

**Fig. 2.** Sample ranked list returned from the query "Mississippian". The map metadata, along with the document metadata, are displayed.

## 4 Automatic Metadata Extraction

The text associated with an image in a document can help interpret the image better. We define two sets of metadata: map level metadata generated from the text accompanying the map, and the host document level metadata.

### 4.1 Map Level Metadata

We consider the following map-level metadata: the caption, the reference text, and the document page containing the map. These three entities can be used for identifying as well as retrieving maps. They are listed as follows, ordered by the likelihood of being useful during a map search.

- *Map caption*.
- *Map reference.* We use the term *map reference* to refer to text in the document that refers to the map. We denote a phrase like "Figure 1" or "Fig 1" in the main body as the *reference point*, and we define the map reference text as the sentence which contains this reference point.
- *Map size.* We use a relative instead of absolute percentage value to represent the map size. Suppose the number of text lines in a page without any figure is $N_T^{nf}$, and the number of text lines in a page with $N_F^f$ figures is $N_T^f$, then the average size of each figure can be computed as $\frac{1-N_T^f/N_T^{nf}}{N_F^f}$.
- *Map figure number and document page number*.

---

[6] http://lucene.apache.org/

First, to extract the caption of a map, we consider each line as a unit and try to identify whether a unit belongs to the classes "caption_begin", "caption_end", or "other". A unit is tagged as caption_begin if all of the following conditions are satisfied: 1) it begins with "Figure", "FIGURE", "Fig", or "FIG" + a number; 2) its font and fontsize are both different from those of the previous unit; and 3) its font size is smaller than that of the document's regular size. Alternatively, a unit is tagged as caption_end if all of the following conditions are satisfied: 1) it ends with "."; 2) its font and fontsize are both different from those of the following unit; 3) its font size is smaller than that of the document's regular size; and 4) all units between the current unit and the most previous caption_begin unit are in the same paragraph and have the same presentation, i.e., the same font and font size. Lines lying between the unit caption_begin to the unit labeled caption_end comprise the caption of a figure.

We define another set of heuristic rules for the reference text extraction. Given a map's figure number and its host document, we first figure out the reference point in the document body by using the following rules: 1) *Key Words*: The reference point is tagged with "Figure", "FIGURE", "Fig", or "FIG" and followed by the figure number of the map; and 2) *Font Size*: The font size of the reference is the same as the regular size of the document. Based on the second rule, we avoid mistaking the map's caption for the map's reference. Once we have acquired the reference point, we extract the sentence containing the reference point as the map's reference text.

### 4.2 Document Level Metadata

Our system extracts document level metadata comprising of the *title*, *abstract*, *author*, *publication date*, and *citations*. In response to users' search queries, the document-level metadata is used to assist in the search (see Section 6). We built learning models to extract document level metadata, as those used in existing digital libraries [17]. The extraction process is not the focus of our paper.

## 5 Map Identification

In this section, we outline an algorithm to identify maps from among other figures using a two-class classifier on a set of features.

### 5.1 Features Extraction

We consider two categories of features for the classifier: basic term-based features and advanced features. First, the algorithm removes stop words, stems words, and eliminates from the vocabulary the terms that appear only once in the whole set of documents. Preprocessing reduces the dimension of the feature vectors by removing terms not helpful in identifying a map.

**Basic Features**. The algorithm uses a set of basic features corresponding to the frequencies of the terms in the text. A *term* is defined to be any alphabetical sequence of characters separated by non-alphabetical characters such as white space. Terms are generated by splitting the text at any non-alphabetical character. The algorithm uses three categories of basic term features extracted from (i) captions, (ii) references, and (iii) a combination of both.

**Advanced Features**. We use the following advanced features:
- *BeginsWith*. This feature is 1 for the first term in a caption and 0 for rest of the terms.
- *FigureNO*. This is one binary feature indicating whether or not the figure's number falls within the range [1,2]. This feature is specific to the archaeology domain. We have observed that usually the first figure of an archeology paper is a map because the paper tends to introduce the geographical information, and then present other related information.
- *LocationNames*. These are two sets of binary features calculating how many words related to locations are used in the figure's caption and reference, respectively. Typically, the text explaining the figure will contain more location-related words if the figure is a map. We use GATE[7] to tag location names in the caption/reference. Next, the number of tagged location names are separated to six intervals: [0], [1,2], [3,5], [5,9], [10, 20], [21, $\infty$]. We define one binary feature for each interval. For instance, if the number of tagged location names falls within a specific interval, the corresponding feature becomes 1 while all other five features remain 0.
- *Size*. This is a binary feature indicating whether or not the image size is larger than $1/3$ of the entire page. We have observed that most large figures in archeology papers are maps because maps bear geographical details, which are usually more complex than figures and require more space for illustrations.

---
[7] http://gate.ac.uk/

### 5.2 SVM Classification

The number of features could be too large for training the model efficiently and labeling other figures. To address this issue, we adopt a statistical method, namely expected entropy loss [18], to select only the features that are effective for the training process.

Expected entropy loss is computed separately for each feature. Then, we rank the features according to their expected entropy loss. Features that are common in both the positive set and the negative set rank lower, while they rank higher if they are effective in discriminating the positive class from the negative one. This approach has been generally adopted for feature selection before training a classifier [19]. We compute the expected entropy loss for each feature and sort all the features by expected entropy loss in descending order. Through setting a threshold, we get feature sets of different sizes. In our experiment, we will test the effectiveness of tuning the size of the feature set.

We consider the map identification process as a classification process that categorizes images into two groups, maps and non-maps. Our system identifies maps using a supervised classifier. First, we train a classifier on the feature vectors generated from the training set in which each image has a class label indicating whether it is a map or not. This classifier is then used to test other images. Cross validation [20] is used for the training and the testing process.

We use Support Vector Machines(SVMs), which are learning machines used for two-group classification problems [21]. SVMs have been proved to be the most applicable tool for text categorization problems because they can handle high dimensional feature vectors, reduce problem caused by over-fitting, and produce solutions that are robust to noise [22]. For our problem, we assume that the classification problem is linearly separable. That is, the set of training data in a high-dimensional space can be classified into two classes by a separating hyperplane. Therefore, to solve the classification problem, it is necessary to construct such an optimal hyperplane in a high-dimensional space.

## 6 An Extended TF-IDF Ranking Model for Maps

Robertson, Zaragoza, and Taylor have proposed a simple extension to BM25 to handle multiple weighted fields [15]. In their scenario, like in our case, a number of fields in one document contribute to the final ranking. The metadata of a map comes from different fields, e.g., the map's caption, reference document title, document abstract, etc. Each field contributes differently to the final ranking and has a different score/weight for its contribution. We give different weights to these $f$ fields as $W_1$, $W_2$, ..., and, $W_f$, respectively such that $W_1 + W_2 + ... + W_f = 1$ without any loss of generality. Given a map $M_j$ with $f$ fields and a query $q$, a straightforward way to combine all fields' contributions is to first take each field as a separate document and compute the cosine similarity between the field and the query, $sim_1$, $sim_2$, ..., and $sim_f$. Then we can combine the weights linearly using these similarities and obtain the similarity score between $M_j$ and $q$: $W_1 sim_1 + W_2 sim_2 + \cdots + W_f sim_f$. However, this is claimed to be problematic by Robertson et. al. [15] because the boost in terms of $TF$ should not be linear, which means the information gained on observing a term at the first time is greater than the information gained on subsequently seeing the same term. It is reasonable to compute weighted $TF$ for each field as discussed in the rest of this section, before the final summation function is applied.

We define a novel scheme named weighted Map Term Frequency and Inverse Map Term Frequency (*MTF-IMTF*) for our map search system. Given a map $M_j$, for a term $T_i$, which appears in the $k$th field $MD_k$ with $W_k$, we define the weighted *MTF* as $W_k \times MTF_{ijk}$, where $MTF_{ijk}$ is the map term frequency of the term $T_i$ in the metadata $MD_k$ of map $M_j$. For the *IMTF*, we use $N_m$ to denote the total number of maps in a collection, and use $n_{ik}$ to denote the number of maps that contain the term $T_i$ in its metadata $MD_k$. Therefore, $\omega_{ijk}$ is computed as $\omega_{ijk} = W_k \times MTF_{ijk} \times \log N_m/n_{ik}$. Similarly, we can get $\omega_{iqk}$ for a query $Q$ if $Q$ specifies a field in which a keyword should occur. Otherwise, if $Q$ consists of only a set of general keywords, the system searches for the keywords in each field, $MD$. In this work, we consider only the latter case, which is more general. The map $M_j$ and the query $Q$ are represented as $D$-dimensional vectors $\overrightarrow{m}_j$ and $\overrightarrow{q'}$, respectively. Here $D = \sum_{i=1}^{f} |MD_i|$. The major difference between the weighted *MTF-MITF* and *TF-IDF* is shown in Figure 3(a) and 3(b). Finally, the similarity between a map $M_j$ and a query $Q$ is computed as the cosine of the angle between $\overrightarrow{m}_j$ and $\overrightarrow{q'}$.

$$sim(M_j, Q) = \cos(\overrightarrow{m}_j, \overrightarrow{q'}) = \frac{\overrightarrow{m}_j \bullet \overrightarrow{q'}}{|\overrightarrow{m}_j| \times |\overrightarrow{q'}|} = \frac{\sum_{i=1}^{D} \omega_{ijk} \cdot \omega_{iqk}}{\sqrt{\sum_{i=1}^{D} \omega_{ijk}^2} \sqrt{\sum_{i=1}^{D} \omega_{iqk}^2}}. \quad (1)$$

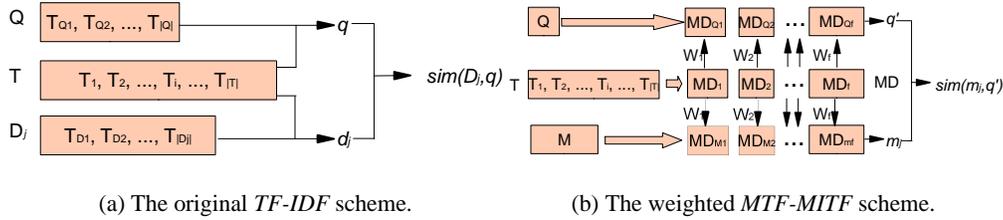

(a) The original *TF-IDF* scheme.  (b) The weighted *MTF-MITF* scheme.

**Fig. 3.** Two *TF-IDF* schemes for map retrieval. Different matadata get different weights when computing the *MTF-MITF*.

Our search engine uses the $MTF - IMTF$ scheme with five sets of metadata: 1) caption, 2) reference text, 3) document title, 4) document abstract, and 5) a set of location names extracted from 1), 2), 3), and 4) above. The weights of these five sets of metadata are denoted as $W_c$, $W_r$, $W_t$, $W_a$, and $W_{loc}$, respectively. By setting and tuning the weights, we obtain various ranking strategies.

Users may be more interested in maps from higher quality documents. To account for this preference, we use several query-independent features of the map($MF$) and its host document($HDF$) to adjust the final ranking. The set $MF$ consists of two factors: 1) the length of map caption $L_c$; and 2) the length of map reference $L_r$. Thus, maps with more detailed explanations get higher ranking. The set $HDF$ consists of four factors: 1) the map frequency $MFreq$; 2) the number of other publications citing the host document $Cite$, which can be obtained from Google Scholar [8]; 3) the venue's prestige $VP$, which can be obtained from some online journal ranking website [9]; and 4) the document freshness $DF$, which is the age of the document measured in years elapsed from the publication year. Boosted by $HDF$, those maps whose host documents contain more maps, have been newly published, appeared in in high quality journals/references, and are cited by more publications can get higher rankings. Finally, normalizing all the above six features, we get the ranking score for the $M_j$ as, $score(M_j, Q) = sim(M_j, Q) \times MF \times HDF$, where $MF = L_c + L_r$, $HDF = MFreq + Cite + VP + DF$.

---

[8] http://scholar.google.com/
[9] http://www.journal-ranking.com/

## 7 Experimental Evaluation

We downloaded a set of archaeology documents from the American Antiquity journal from JSTOR[10]. Table 1 shows some basic statistics for our collection.

| Items | Number |
|---|---|
| Papers | 2000 |
| Papers with Fig. | 465 |
| Fig. | 2090 |
| Papers with Maps | 278 |
| Maps | 536 |

**Table 1.** Statistics of the collection used in the tests.

| Field | $P(\%)$ | $R(\%)$ |
|---|---|---|
| Caption | 100 | 97.7 |
| Reference | 96.3 | 94.8 |

**Table 2.** Performance of map caption and reference extraction.

We use four metrics, *precision(P)*, *recall(R)*, *accuracy(A)*, and *F-measure(F)*. Let $W$ be the number of true positives predicted as positive, $X$ as the number of true positives predicted as negative, $Y$ as the number of true negatives predicted as positive, and $Z$ as the number of true negatives predicted as negative. The above four metrics can be computed as: $P = \frac{W}{W+Y}, R = \frac{W}{W+X}, A = \frac{W+Z}{W+X+Y+Z}$, and $F = \frac{2*P*R}{P+R}$.

### 7.1 Metadata Extraction

We applied our metadata extraction heuristic rules on 2000 archeology papers to extract 2090 figure metadata and verified the results manually. Table 2 shows that all the extracted captions are correct while there are 2.3% true positive captions missed. The system missed some captions because in some old publications, figure captions are embedded in their images instead of being included in text. Our experiment missed those captions because we do not explore the content inside the figure. For reference extraction, both $P$ and $R$ are above 90%. Some errors are very difficult to resolve. For example, a reference in one paper points to a figure in another paper. Without deep semantic understanding, this reference can be mistaken as the reference to a map in the current document. Some positives were missed because of the quality of the transformed result in the OCR system. For example, the OCR system recognizes "Figure 1b" wrongly as "Figure 16".

### 7.2 Map Identification

We used SVM-Light [11] on our dataset of 2090 figures.

**Effectiveness of Feature Selection**. We partitioned the features into different categories and ran the SVM classifier on each category. These categories include: 1) *Caption*: All terms in the caption; 2) *Reference*: All terms in the reference; 3) *Caption+Reference*: All terms in both caption and reference; 4) *All features*: All terms in both caption and reference, as well as other advanced features (see Section 5); and 5) *Single feature*: Whether the caption begins with "map" or not. In these experiments, the default parameter setting, i.e., linear kernel, in SVM-Light were used. Based on the expected entropy, we sorted the features in each feature set and tuned the size of each feature to see its impact to the classifier's performance.

Figure 4 shows the result of comparing the above strategies using five-fold cross validation. As expected, the SVM method shows significant improvement over the rule-based single feature method. The rule-based method has the highest $P$ as 1, which means that each figure whose caption begins with "map" is a real map. However, the recall of this method is the lowest (0.231). Second, we find that the terms in captions are more important than those in references for map identification. The caption-based classifier significantly outperforms reference-based classifier in terms of all metrics. In addition, combining both the basic and advanced features results in the best performance among all the methods. This clearly shows the effectiveness of the advanced features we discussed in Section 5. Finally, tuning the size of features has a huge influence on the classification performance. Figure 4(a)-4(d) illustrate that the value of each metric improves when the number of features becomes larger. Particularly, the value grows sharply at the beginning. After the feature number reaches 5,000, the value either decreases or remains stable. Regarding the computation time, Figure 4(e) and 4(f) show that both the training and the testing time go up rapidly when the feature size increases. The gradient is much sharper when the feature size becomes larger. Therefore, we conclude that for our dataset the optimal setting for the feature size is 5,000. In the following experiments, we fix the feature set as "all features" and the feature size as 5,000.

---
[10] http://www.jstor.org/
[11] http://svmlight.joachims.org/

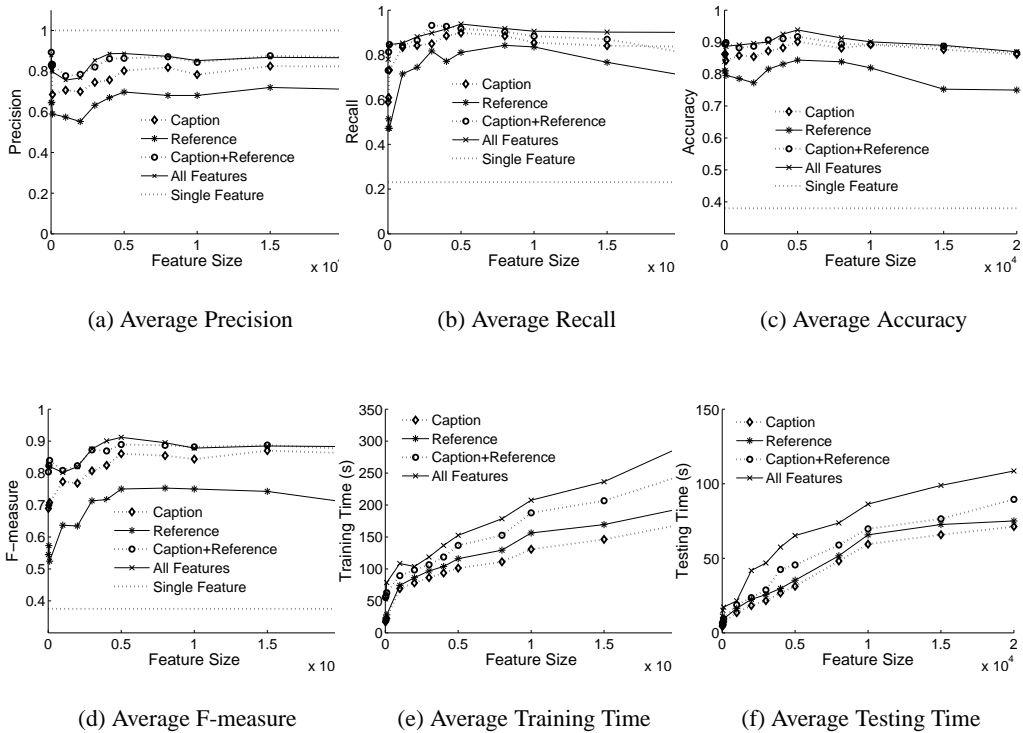

**Fig. 4.** Map identification using different settings of feature sets.

To further study the effectiveness of each feature set, we take a close look at the top 10 features with the highest expected entropy for each feature set, as listed in Table 3. In this table, a minus sign indicates the negative terms for the class of maps, which means these tokens more frequently appear in the non-maps class than in the maps class. A capitalized beginning in a word indicates that this token appears at the beginning of the caption. All these features are expected to best describe the class of maps.

**Impact of Different SVM Kernels**. Based on the same feature set, we experimented with three different types of kernels for the SVM classifier. These are linear, polynomial, and Gaussian kernels. Table 4 shows the comparison result when using these kernels. We can see that SVM linear outperforms the other methods in terms of average *F-measure*. To further evaluate the effectiveness of these kernels, we apply a *two-sample hypothesis testing about means*, which is a commonly used statistical method in large sample tests [23]. We present the numerical results of the t-values in Table 5. The results show that the t-value of comparison between the linear kernel and the polynomial kernel is larger than the critical value (at the confidence level of 0.05) while the other t-values are smaller than the critical value. Our results imply that, with 97.5% confidence, SVM linear significantly outperforms SVM poly, while the difference between other methods is not statistically significant.

| Feature sets | Top 10 features |
|---|---|
| Caption | map, location, site, show, contour, -text, mention, topography, area, -plot |
| Reference | locate, km, river, site, area, north, map, site, valley, northern |
| Caption & Reference | map, location, site, area, km, river, north, show, valley, region |
| All features | map, location, Map, Location, Site, site, area, north, -Plot, river |

**Table 3.** The top 10 features for each feature set used for the SVM classifier; all these features are expected to best describe the class of maps.

| Approaches | $P(\%)$ | $R(\%)$ | $A(\%)$ | $F(\%)$ |
|---|---|---|---|---|
| Begins with "map" | 100 | 23.1 | 38.45 | 37.53 |
| SVM (linear) | 88.7 | 91.6 | 92.96 | 90.13 |
| SVM (poly) | 86.1 | 89.2 | 89.22 | 87.62 |
| SVM (Gaussian) | 87.3 | 90.1 | 90.13 | 88.68 |

**Table 4.** Map identification performance using different kernels for SVM classifier.

| Comparison between kernels | t-values |
|---|---|
| linear VS poly | 1.8734 |
| linear VS Gaussian | 1.3281 |
| Gaussian VS poly | 0.8377 |

**Table 5.** The $t$-values for pairs of *F-measure* samples. SVM linear outperforms SVM poly at the significance level of 0.05.

### 7.3 Map Ranking

We used 25 keyword-based queries to search 536 maps from 278 papers. The queries are grouped into five categories: 1) *General locations*: United States, Mississippian, Virginia, Mexico, Columbia; 2) *Site names*: Hopewell, Clovis, Cahokia, Mesa Verde, Moundville; 3) *General objects*: pottery, stone, glass, kernel, bone; 4) *Archeological objects*: shell, hammer, bowl fragment, playa; and 5) *Combined queries*: stone Mexico, pottery Hopewell, shell Mississippian, glass Mesa Verde, kernel Virginia. The average query length is 1.32 words.

We ran our ranking methods for each query and put the 30 highest ranked maps in each returned ranking into a pool for evaluation. The maps in the pool were then manually tagged as relevant or non-relevant with respect to each query. As a result, for each query, we have a pool of maps, labeled as relevant or non-relevant. This pool method avoids the need to evaluate the whole collection which could be extremely labor-intensive.

We computed average $P$ for all queries at each of the 11 $R$ levels, 0%, 10%, ..., and 100%. $P$ at 0% $R$ is the precision when the first relevant document is seen in the returned list, starting from the top. $P$ at 100% $R$ is the precision when all the relevant documents have been seen in the ranked list, starting from the top. Figure 5 shows the *P-R* figures for various indexing and ranking functions.

First, we implemented two ranking variances each of which is based on a single source, caption only, and reference only. Two other baselines are also implemented for comparison: full text indexing and Customized Google Desktop search engine, which is a Google Desktop API indexes only the directory containing the 278 PDF files. For the baseline methods, we input the queries with an additional keyword, "map". However, their returned results are not maps but entire documents. We take out the "highlighted text" for each result, which the search engine's ranking strategy considers to be the most relevant to the test query. Then, the page in the document which contains the "highlighted text" is located. The map on that page is considered to be the result. If there is no map on that page, a miss is counted. We show the results in Figure 5(a). The two baseline methods both have the worst performance because they do not take any efforts to deal with maps. The other two strategies can both locate maps in the documents. The caption-based method achieves higher precision than reference-based one before the 80% $R$, because a map's caption concentrates on the map's content while the refeence may be referring to other content in the document. In the following experiments with multiple sources, we will set higher weights to the terms in captions.

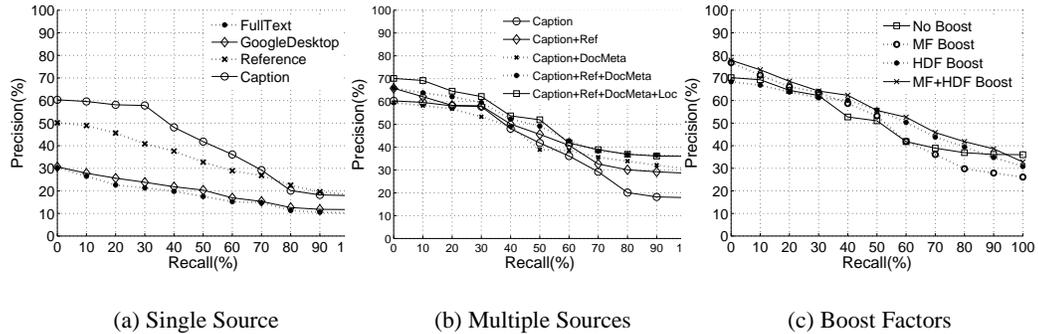

(a) Single Source  (b) Multiple Sources  (c) Boost Factors

**Fig. 5.** The average precision-recall figures for 25 test queries.

We ran four combinations to test the effectiveness of multiple sources: (1) Caption+Reference, where $W_c = 0.66, W_r = 0.33, W_t = W_a = W_{loc} = 0$; (2) Caption+DocMeta, where $W_c = 0.6, W_t = 0.2, W_a = 0.2, W_r = W_{loc} = 0$; and (3) Caption+Reference+DocMeta, where $W_c = 0.5, W_r = 0.3, W_t = 0.1, W_a = 0.1, W_{loc} = 0$; and (4) Caption+Ref+DocMeta+Loc, where $W_c = 0.5, W_r = 0.2, W_{loc} = 0.1, W_t = 0.1, W_a = 0.1, MF = HDF = 1$. Note that here we do not consider the boosting factors, which means we set $MF = HDF = 1$. The values of weights are set according to the previous experimental results. Figure 5(b) illustrates the comparison results. The ranking method that uses both terms in captions and references is more precise than caption-based ranking. After including the document metadata, strategies (2) and (3) above both yield slightly better retrieval results than usng strategy (1) . Thus, the document's title and abstract give some useful information for map search. Furthermore, the inclusion of location keywords in test (4) leads to an increase in the average precision.

Finally, we present an evaluation of the impact of the boosting factors (i.e., *MF* and *HDF*) in Figure 5(c). It shows that including the *HDF* boost factor could improve the average precision. This is because, usually, the users are more interested in maps extracted from popular and important host documents than unpopular and unimportant ones. The highest average precision is obtained when both boost factors are included. Particularly, the improvement becomes more significant when the recall is higher. This indicates that the boost factors greatly contribute to improving the quality of the top ranked results. This is especially helpful to a Web digital library system, where precision is very important among the top ranked documents.

The above test results indicate that, first, the map metadata set we proposed is very useful to understand the map's content; and second, the weighted *MTF-IMTF* scheme can accurately predict the relevance of a map to a query.

## 8 Conclusions and Future Work

We have proposed a novel system that enables searching for maps in digital academic documents. We select a set of features that help identify maps. Our system first extracts figures from documents and classifies them as maps and non-maps. It identifies the caption and references of the maps and indexes them. We have proposed a ranking function that utilizes the captions and references of the maps, as well as document-level features like the document title, document age, document popularity and importance of publication venues to enable map search. Experimental results performed on JSTOR archaeology journal document show the effectiveness of our chosen features and ranking methods.

In the future, there are several interesting directions to pursue. First, we plan to include image analysis tools for indexing and utilizing text appearing inside the maps themselves. Second, we realize that different classification algorithms may have different impacts on map identification. We will investigate other classifiers besides SVM and find the most suitable for our application. Finally, we will look into location keywords and take into account spatial queries in the map retrieval system.